\documentclass[12pt,preprint]{aastex}

\usepackage{epsfig}
\usepackage{xcolor}

\begin{document}

\title{Limits on the Mass and Abundance of Primordial Black Holes from Quasar Gravitational Microlensing}

\author{E. MEDIAVILLA\altaffilmark{1,2}, J. JIM\'ENEZ-VICENTE\altaffilmark{3,4}, J. A. MU\~NOZ\altaffilmark{5,6}, H.  VIVES-ARIAS\altaffilmark{5},  J. CALDER\'ON-INFANTE\altaffilmark{3}}

\altaffiltext{1}{Instituto de Astrof\'{\i}sica de Canarias, V\'{\i}a L\'actea S/N, La Laguna 38200, Tenerife, Spain}
\altaffiltext{2}{Departamento de Astrof\'{\i}sica, Universidad de la Laguna, La Laguna 38200, Tenerife, Spain}
\altaffiltext{3}{Departamento de F\'{\i}sica Te\'orica y del Cosmos, Universidad de Granada, Campus de Fuentenueva, 18071 Granada, Spain}
\altaffiltext{4}{Instituto Carlos I de F\'{\i}sica Te\'orica y Computacional, Universidad de Granada, 18071 Granada, Spain}
\altaffiltext{5}{Departamento de Astronom\'{\i}a y Astrof\'{\i}sica, Universidad de Valencia, 46100 Burjassot, Valencia, Spain.}
\altaffiltext{6}{Observatorio Astron\'omico, Universidad de Valencia, E-46980 Paterna, Valencia, Spain}

\begin{abstract}

The idea that dark matter can be made of intermediate-mass primordial black holes in the  
$10M_\sun \lesssim M \lesssim 200M_\sun$ range has recently been reconsidered, particularly 
in the light of the detection of gravitational waves by the LIGO experiment. The existence 
of  even a small fraction of dark matter in black holes should nevertheless result in noticeable 
quasar gravitational microlensing.
Quasar microlensing is sensitive to any type of compact objects in the lens galaxy, to their
abundance, and to their mass. We have analyzed optical and X-ray microlensing data from 24 gravitationally 
lensed quasars to estimate the abundance of compact objects in a very wide range of masses. We
conclude that the fraction of mass in black holes or any type of compact objects is negligible 
outside of the $0.05 M_\odot \lesssim M \lesssim 0.45 M_\odot$ mass range {and that it} 
amounts to $20 \pm5$\% of the total matter,  in agreement with the expected
masses and abundances of the stellar component. 
Consequently, {the existence of a significant population of intermediate-mass primordial black holes appears to be  inconsistent with current microlensing observations. Therefore, }
primordial massive black holes are a very unlikely source of the gravitational radiation detected by LIGO. 

\end{abstract}


\section{Introduction \label{intro}}

The lack of evidence for elementary particles that could explain dark matter 
(see Frampton, 2016, Chapline \& Frampton, 2016, and references therein) and the 
detection of gravitational waves by the LIGO experiment produced by black holes 
(Abbott et al. 2016), have recently renewed  interest in the  possibility that
dark matter could consist of primordial black holes (hereafter PBH) in the 
intermediate-mass range $1 M_\sun < M < 1000 M_\sun$ (Carr et al. 2016). Lower masses 
are excluded by galactic microlensing experiments (e.g. Alcock et al. 1998, 
Tisserand et al. 2007), and larger masses (which are not constrained by galactic 
microlensing) are assumed to disrupt stellar systems (e.g. Brandt 2016). Irrespective 
of the precision of these constraints (Hawkins 2015, Carr et al. 2016, Bird et al. 2016, 
Green 2016) and even of the existence of PBH, we would like to draw attention to the fact that the more 
universal\footnote{It is not the aim of this paper to discuss the limits on 
compact bodies set by galactic microlensing experiments, but they are dependent on the 
modeling of the Galactic halo and subject to controversy (see, e.g., Hawkins 2015).} 
and potentially very powerful scenario of quasar microlensing has  scarcely been
analyzed in this context. In fact, quasar microlensing can be used to independently 
explore a very wide mass range (Dalcanton et al. 1994) and is sensitive to the 
abundance and masses not only of intermediate mass BH but of any population of 
compact objects such as small mass BH, planets, normal or exotic stars  (Mazur
\& Mottola 2004, Chapline 2005), and stellar remnants. 

Quasar microlensing manifests itself in the variation of the brightness of  images of
multiply imaged quasars induced by compact objects in  the lens galaxy (Chang \& Refsdal 1979, 
Wambsganss 2006). It provides a natural extension of galactic microlensing experiments 
in the search of MACHOs to the extragalactic domain (Mediavilla et al. 2009). Using 
single-epoch flux measurements of the images of samples of lensed quasars, several authors 
(Schechter \& Wambsganss 2004, Mediavilla et al. 2009, Pooley et al. 2012, Schechter et
al. 2014, Jim\'enez-Vicente et al. 2015a,b) have studied the abundance of compact objects
in  lens galaxies. However, these studies  emphasize  the determination of the 
fraction of mass in compact objects by considering a limited range of masses, preferably close 
to typical stellar values. Likely for this reason, they have not been taken into account in 
the discussion of PBH abundance (Carr et al. 2016).
Our proposal here is to use  available quasar microlensing magnification measurements 
to constrain the fraction of dark matter in compact objects, but considering now as wide a mass 
range as possible.

\section{Dependence of the Abundance of Compact Objects on their Masses from Quasar Microlensing Data\label{pdf}}

In principle, it could be considered that the compact objects acting like microlenses are 
distributed according to a given mass spectrum. For finite size sources,
 microlensing is sensitive to the mean mass of the microlenses so that the larger the masses, 
 the broader the histogram of microlensing magnifications. 
However, magnification statistics are barely sensitive to the slope of the mass function of 
the microlenses (Congdon et al. 2007) and are, in practice,  degenerate with the single-mass
case\footnote{The degeneracy may be broken when a similar contribution to the mass density 
from microlenses at opposite ends of the mass spectrum exists { (bimodality) and the type of lensed image  maximizes the effects of bimodality (saddle-point images of high magnification, see Schechter, Wambsganss \& Lewis 2004)}.} (Wyithe \& Turner, 2001; 
Schechter, Wambsganss \& Lewis 2004; Congdon et al. 2007).
We will adopt here, then, { a single-mass distribution of compact objects of identical mass plus a smooth distribution of dark matter representing the rest of the mass. Notice that this composite mass function can also reproduce some basic features of the bimodal mass distribution with two single-mass populations of compact objects of} very different masses but comparable contributions to the mass density  { (see Schechter, Wambsganss \& Lewis 2004, and also Gil-Merino \& Lewis 2006). We will discuss this issue in more detail later}.

Our study is based on the data and simulations from Jim\'enez-Vicente et al. (2015a) in the 
optical range, and Jim\'enez-Vicente et al. (2015b) in X-rays{, } who used microlensing magnification
measurements from 24 gravitationally lensed quasars. In these studies an estimate of the abundance
of microlenses (i.e. of compact objects) is obtained by  comparing the observed 
microlensing magnification for each pair of lensed quasar images, $\Delta m_{ij}$, with the simulated
values for these measurements as a function of the physical parameters of interest. We refer the 
reader to those works for  details on the microlensing datasets, simulations, and statistical 
procedure. The most relevant physical parameters affecting the amount of observed microlensing are
the fraction of total matter in microlenses, $\alpha$,  their average mass, $M$, and the size of the accretion disk source, $r_s$. 
For the present study, focussed on the abundance and mass of the microlenses, we will consider 
$\alpha$ and $M$ as the physical variables and $r_s$ as a fixed parameter { which, to avoid circularity, we will relate to reverberation mapping studies. The impact of the uncertainties on this parameter,} nevertheless, 
will be discussed later. Following a similar procedure as in Jim\'enez-Vicente et al. (2015,a,b),
we can derive from the simulations the probability of the observed  microlensing measurements,
$\Delta m_{ij}$, conditioned to the parameters  ($\alpha,M$), $p(\Delta m_{ij}|\alpha,M)$, and, 
using Bayes's theorem, the probability density function {(PDF)} for the parameters 
$L(\alpha,M|\Delta m_{ij}) \propto p(\Delta m_{ij}|\alpha,M)$.

{ The natural scale of microlensing is the Einstein radius which increases with the microlenses mass, 
$\eta \propto \sqrt{M}$, and  corresponds to the size of the region in which the 
magnification induced by  an isolated microlens is noticeable. As a consequence of this, gravitational 
microlensing is invariant under a transformation of the mass of the microlenses, $M\to M'$, if the size of 
the source is transformed as $r_s\to r_s\sqrt{M'/M}$ (mass-size degeneracy). Using this invariance, it is
straightforward to transform the {PDFs} $L(r_s,\alpha|\Delta m_{ij})$
obtained by Jim\'enez-Vicente et al. (2015a,b) for a fixed $M$ to $L(M,\alpha|\Delta m_{ij})$ 
adopting now a fixed value for  $r_s$. }

{ There are two main sources of $r_s$ measurements: microlensing of quasars with central super-massive BH masses in the ${\cal M}_{BH}\sim 10^8-10^9 M_\odot$ range, and reverberation mapping (hereafter RM; Blandford \& McKee 1982, Peterson 1993), that has  provided sizes for several nearby AGN with ${\cal M}_{BH}\lesssim 10^8 M_\odot$. To break the mass-size degeneracy of microlensing we can compare the results of both methods in the common range, anchoring the $r_s({\cal M}_{BH})$ dependence inferred from microlensing (Morgan et al. 2010, Mosquera et al. 2013) to RM measurements. When this comparison is done in the optical (Edelson et al. 2015, Fausnaugh et al. 2016 and references therein) it is found that microlensing size estimates (Mosquera et al. 2013)  agree remarkably well 
with those inferred from RM. Thus, we 
can adopt, from the $r_s({\cal M}_{BH})$ dependence (Mosquera et al. 2013) validated with RM, a reference value for $r_s$ of $5\,$lt-day corresponding to ${\cal M}_{BH}\sim  10^9 M_\odot$. Other possible prescriptions also based in RM results would produce very similar values for the size of the accretion disk\footnote{ The fit of the combined RM and microlensing data by Edelson et al. (2015) for ${\cal M}_{BH}\sim 10^9 M_\odot$, gives $r_s=6.5\,$lt-day. An extrapolation using the theoretical $r_s\propto {\cal M}_{BH}^{2/3}$ law to the $10^9 M_\odot$ mass of the results obtained from RM of three well studied cases, MGC 08-11-011, NGC 2617 and NGC 5548 (Fausnaugh et al. 2016) results in $r_s=4.3\,$lt-day. { See also Jiang et al. (2016).}}.

In the case of X-ray observations the results and their interpretation are less clear, but available RM measurements of many AGN (see Kara et al. 2016, Uttley et al. 2014 and references therein) are in reasonable agreement with microlensing studies with a size proportional to the central black-hole mass of roughly 10 gravitational radii, that for ${\cal M}_{BH} \sim 10^9 M_\odot$ corresponds to 0.6 lt-day. In any case,}  the dependence on { size} can be made explicit as a scaling factor proportional to $r_s^2$ (see below).

The resulting  {PDFs}, $L(M,\alpha|\Delta m_{ij})$, 
are shown in Figure 1 for the optical microlensing measurements analyzed in Jim\'enez-Vicente et al.
(2015a) {corresponding to a rest wavelength of $\sim$1736\AA} and in Figure 2 for the X-ray data discussed in Jim\'enez-Vicente et al. (2015b). 
The first result (see Figure 1) is that the optical data strongly constrain the compact object 
masses to the range $0.05M_\odot(r^{opt}_s/ 5 \,{\rm lt-day})^2 \lesssim M \lesssim0.45M_\odot(r^{opt}_s/ 5 \,{\rm lt-day})^2$ 
at the  90\% confidence level. In principle, a change in $r^{opt}_s$ toward larger values
could push up the limits of the mass interval, but notice that the accretion disk size 
needed to include microlenses of $\sim10M_\odot$ at the { upper limit of the} 90\% significance level interval 
is $r^{opt}_s\sim 24\,\rm$lt-day, which is very difficult to accept { taking into account the RM estimates for the size of the accretion disks  (see Edelson et al. 2015, Fausnaugh et al. 2016 and references therein)}. The X-ray data also
concentrate the probability on relatively low masses but allow a relatively unconstrained 
upper limit, with  a confidence interval of 
$0.04M_\odot(r^X_s/ 0.6 \,{\rm lt-days})^2 \lesssim M \lesssim 35M_\odot(r^X_s/ 0.6 \,{\rm lt-days})^2$ at 
the 68\% significance level. Also in this case, larger values of the source size could extend the
upper limit of the allowed mass interval, but sizes significantly greater than $0.6\, \rm$lt-day
are not expected { according to RM results (see Kara et al. 2016, Uttley et al. 2014 and references therein)}. The low sensitivity of X-ray observations to the microlens mass for
$M\ge 1M_\odot$ reflects the fact that the X-ray source becomes point-like (and hence 
insensitive to the mass function\footnote{ For a point source the microlensing magnification
probability distribution is only weakly sensitive to any property of the mass function of 
the microlenses, except for extreme bimodal cases (see Schechter, Wambsganss \& Lewis 2004 
and references therein).}) with respect to the natural scale of
microlensing { (the Einstein radius, $\eta \propto \sqrt{M}$)} in the upper region of the mass range of interest.

The second result, already discussed in the stellar mass range {\bf(}see Mediavilla et al. 2009, Schechter et al. 2014,
Jim\'enez-Vicente et al. 2015a,b), is that the estimated
abundance of microlenses, $\alpha=0.2\pm 0.05$  (see Figures 1 and 2), is in good agreement
with the expected abundance of stars in the lens galaxies (Jim\'enez-Vicente et al. 2015a,b), 
leaving barely any room for an extra population of compact objects.  { We can wonder to what extent is this result dependent on the assumption that the microlenses mass distribution is degenerate with the single-mass case. Specifically,  could an additional population of massive BHs, providing a fraction of mass comparable to the one supplied by the stars (bimodal distribution), have passed unnoticed in our previous analysis? Only a few lensed quasar images in our sample are prone to show the effects of bimodality, and, even in this case (see Appendix A),} { we find that the statistical microlensing properties of a bimodal mass distribution, including normal stars and a significant population of intermediate mass black holes,  can only be reproduced by a single-mass analysis if the average mass is increased by an order of magnitude with respect to the typical stellar mass.}
 {Thus, in so far as we have recovered a rather low mass in our single-mass analysis, the existence of a significant population of massive BH can be discarded} { even for a bimodal distribution, an extreme (but plausible) case of non-smooth mass spectrum.}

A final discussion about the upper limit of the allowed mass range is pertinent. 
In our optical study, which determines the upper mass limit,  the microlensing magnification 
of the quasar continuum is defined with respect to the mean magnification (i.e. the magnification 
in the absence of microlensing). This is experimentally determined from the ratio of the emission 
line cores which are generated in a region much larger than the accretion disk size (see, e.g., { Zu et al. 2011}, Guerras et al. 2013) and, hence, 
less affected by microlensing (which is size sensitive so that the larger the size, the smaller 
the microlensing magnification).  On the other hand, the effects of microlensing are noticeable 
in a region with a size similar to the Einstein radius, $\eta$, proportional to $\sqrt{M}$. Thus, 
we expect that, below a threshold mass, $M_{opt}$, microlensing will affect the tiny region 
generating the continuum but not the far larger emission line region. Above $M_{opt}$ 
microlensing may affect both the continuum and emission lines, and we would need to consider another
way to define the mean magnification baseline. As a conservative lower limit to the size of the 
region generating the core of the emission lines, we can take the average size of the broad line
region. For the typical luminosity of a lensed quasar, $L=10^{44.5}\,\rm erg/s$ (Mosquera \& 
Kochanek 2011), this size corresponds to a half-light radius of $R_{1/2}=94\,\rm$lt-day (according 
to the best-fit correlation induced from reverberation mapping results, Zu et al. 2011). This size 
corresponds\footnote{For a typical lens system with source and lens redshifts of 2 and 0.5, 
respectively.} to the Einstein radius of a $M_{opt}\simeq 22 M_\odot$ microlens that we can 
take as the mass threshold. To go above this value we can take as reference the mid-infrared
emission that arises from a significantly larger region (several parsecs in size, Burtscher 
et al. 2013) than the optical. Although there are only a few observations in the mid-infrared, 
they confirm (see Mediavilla et al. 2009) the mean magnifications obtained in the optical 
using the emission lines cores and allow us to extend the upper limit to $M_{IR}\ge 10^5 M_\odot$.
For quadruple systems, the mean magnification baseline can also be determined by modeling the lens 
system (independently of any mass threshold) and  analysis of the microlensing data leads to a 
low abundance in compact objects compatible with the expected contribution of the stars (Pooley et al. 2012, Schechter et al. 2014). 

To study the possible abundance of the less compact and very much more massive dark matter 
clumps (generating millilensing instead of microlensing), Dalal \& Kochanek (2002) also
studied several quadruple lenses{, finding} a relatively low abundance of clumps ($\alpha_{clumps}<0.07$ at
the 90\% confidence; see also Vegetti et al. 2014 and Vives-Arias et al. 2016). Given that{, according to the pseudo-Jaffe model (Mu{\~n}oz et al. 2001) used in the previously cited works}, clumps 
are less efficient in generating millilensing effects than point objects of the same mass,
even lower abundances are expected in the latter case. Thus, no matter how large the masses of
a supposed population of compact objects are, their microlensing effects on quasars should have been noticed.

\section{Conclusions}

We have used optical and X-ray quasar microlensing data to study the dependence of the abundance of 
BH and, in general, of any population of compact objects in the lens galaxy on their mass obtaining the following results:

1 - A $20 \pm 5$\% of the total mass in the lens galaxies can be in compact objects with masses 
in the $0.05M_\odot(r^{opt}_s/ 5 \,{\rm lt-day})^2$  $\lesssim M \lesssim0.45M_\odot(r^{opt}_s/ 5 \,{\rm lt-day})^2$ 
range {(90\% confidence interval)}, in agreement with the expected masses and abundance of the stellar component. This result was obtained for 
a single-mass  population but holds for  a population distributed according to a reasonably smooth mass spectrum with the same  mean mass. 

2 - {To probe the dependence of this result on the adopted mass function,  we have also considered an extreme case of non-smooth mass spectrum, a bimodal  distribution of stars and BHs, finding that} the existence of a population of { intermediate mass} compact objects with weight comparable to that of the stars in the mass density is also discarded. {This result was obtained for a highly magnified lensed quasar image. Although a thorough treatment could explore more generic mass functions and other type of lensed quasar images, the adopted hypotheses are conservatively chosen to maximize the effect of the departures from a smooth mass function (Schechter, Wambsganss \& Lewis 2004).} 

3 - These results {are incompatible with} the existence of any significant population of massive black holes
and therefore render very unlikely the generation of the gravitational waves detected by LIGO by
{ PBH}, as this hypothesis requires  a large fraction of the dark matter to be constituted by { PBH} (Bird et al. 2016).

\acknowledgements{We thank the referee for the thorough revision of the paper. We thank C.S. Kochanek and P. Schechter for helpful suggestions. This research was supported by the Spanish MINECO with the grants AYA2013-47744-C3-3-P and AYA2013-47744-C3-1-P. JAM is also supported by the Generalitat Valenciana with the grant PROMETEO/2014/60. JJV is supported by the project AYA2014-53506-P financed by the Spanish Ministerio de Econom\'\i a y Competividad and by the Fondo Europeo de Desarrollo Regional (FEDER), and by project FQM-108 financed by Junta de Andaluc\'\i a.}

\appendix

\section{{ Single-mass analysis of microlensing statistics arising from a bimodal mass spectrum.}}

{ In this Appendix we discuss to what extent are the results of our analysis dependent on the assumption that the microlenses mass distribution is degenerate with the single-mass case. This assumption holds for any smooth mass function and there is agreement (Wyithe \& Turner, 2001; 
Schechter, Wambsganss \& Lewis 2004; Congdon et al. 2007) in that only markedly bimodal distributions with a large and comparable contribution to the mass density from microlenses of very different masses may break the degeneracy, and only for images that maximize the effects of this bimodality (saddle-point images of high magnification). Only a few of the images in the samples considered by  Jim\'enez-Vicente et al. (2015a,b) fulfill this last condition (less than 10\% (20\%) saddle-point images with magnifications greater than 10 (5)). 

Nevertheless, in order to analyze the impact of bimodality in our previous single-mass based  analysis,  we have considered, conservatively, the case  of an image with convergence and shear $\kappa=\gamma=0.55$ which strongly favors the effects of bimodality (Schechter, Wambsganss \& Lewis 2004). To do this study, we simulate microlensing of a 5 lt-day source by a bimodal mass function of compact objects including stars ($M_*\sim 0.3M_\odot$) and a population of BH ($M_{BH}\sim 30 M_\odot$) with equal  contributions to the fraction of mass in compact objects ($\alpha_*=\alpha_{BH}=0.2$).}
{Then, the resulting  {PDF} of microlensing magnifications is compared with different  {PDFs}  obtained for the single-mass model in a range of values of the mass fraction, $\alpha$, and the mass, $M$, of the compact objects. We use a  $\chi^2$ test to assess the similarity between  {PDFs}.}
{ In Figure 3 we represent the} {reduced} 
$\chi^2(\alpha,M)$. The best fits correspond to  $0.15 \lesssim \alpha \lesssim 0.3$, and, $2 M_\odot \lesssim M \lesssim 3.5 M_\odot$ ($\pm1\sigma$ {confidence intervals}).  On the other hand,  we confirm that the likelihood of any  model with $M\lesssim 0.5M_\odot$, that could represent the stellar component would have been very low if a bimodal distribution including intermediate mass BH existed. In Figure 4 we represent the probability of microlensing magnifications for the bimodal mass spectrum, the single-mass model typical of the stellar population  ($\alpha=0.2$, $M= 0.3M_\odot$), and the one corresponding to the best fit ($\alpha=0.2$, $M= 2 M_\odot$), to explicitly show the very poor matching of the $0.3M_\odot$ model to the bimodal distribution. 

Thus, we can conclude that (even if most of our images were of the type that maximizes the effects of bimodality, which is not the case) an additional population of massive BH providing a fraction of mass comparable to the one supplied by the stars, would have shown up  in our single-mass analysis by shifting (by about one order of magnitude) the estimated mass towards higher values, while we have recovered the typical masses of normal stellar populations instead.  {This conclusion has been obtained for a quasar image with $\kappa=\gamma=0.55$ {and for a bimodal distribution of microlenses}. These are two reasonable and conservative assumptions that strongly simplify the calculations of a more comprehensive approach which may explore the ($\kappa,\gamma$) plane  and consider more generic mass functions.}

{ An additional result of} {our} {analysis is that the single-mass plus smooth dark matter model has provided a rather good approximation to the magnification statistics in an extreme bimodal case. This is not unexpected as the smooth distribution of matter can, to some extent, reproduce the effect of the population with smaller mass (see Schechter, Wambsganss \& Lewis 2004, Gil-Merino \& Lewis 2006). }

\clearpage

\begin{figure}[h]
\vskip -1 truecm
\plotone{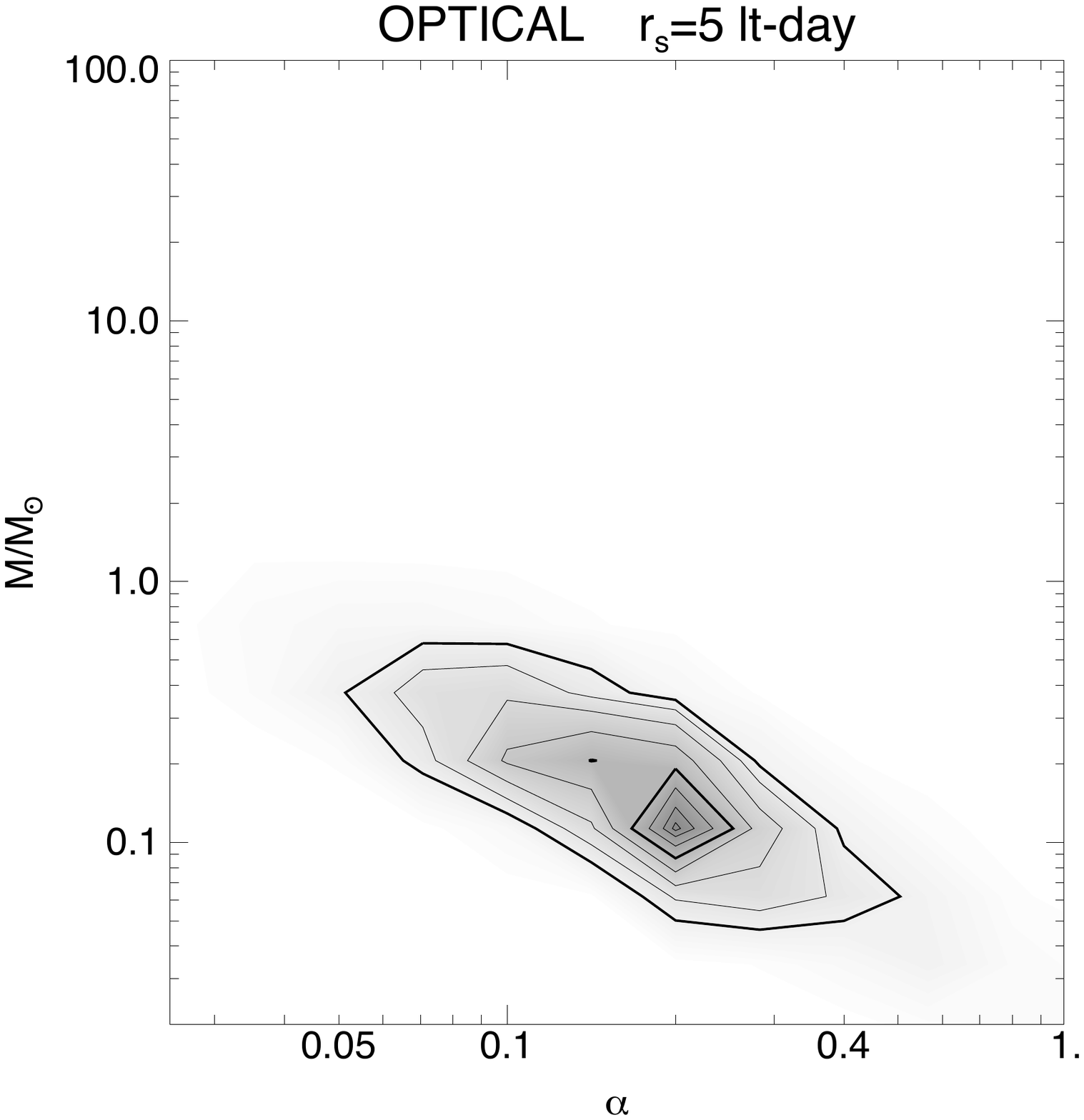}
\vskip -2 truecm
\caption{Likelihood function, $L(\alpha,M|\Delta m_{ij})$,  for the BH (or any type of 
compact objects acting like microlenses) mass fraction, $\alpha$, and mass, ${M}$, obtained 
from microlensing optical data considering a $r_s=5\,$lt-days size source. The contours are
drawn at likelihood intervals of $0.25\sigma$ for one parameter from the maximum. The contours at $1\sigma$ and $2\sigma$ are heavier.}
\end{figure}
\begin{figure}[h]
\vskip -2 truecm
\plotone{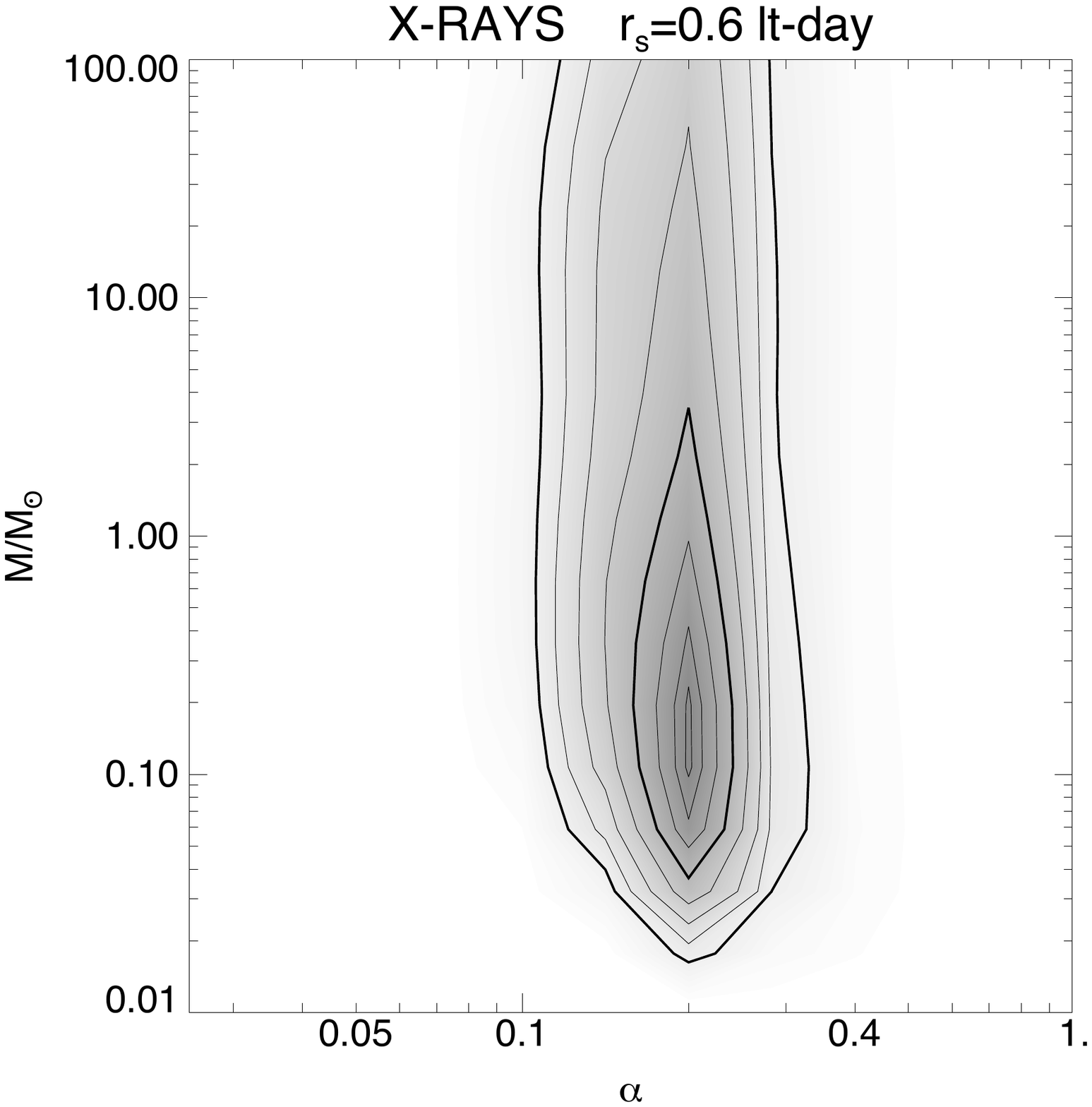}
\vskip -1 truecm
\caption{Likelihood function, $L(\alpha,M|\Delta m_{ij})$,  for the BH (or any type of
compact objects acting like microlenses) mass fraction, $\alpha$, and mass, ${M}$, obtained
from microlensing X-ray data considering a $r_s=0.6\,$lt-days size source. The contours are 
drawn at likelihood intervals of $0.25\sigma$ for one parameter from the maximum. The contours at $1\sigma$ and $2\sigma$ are heavier.}
\end{figure}
\begin{figure}[h]
\vskip -2 truecm
\plotone{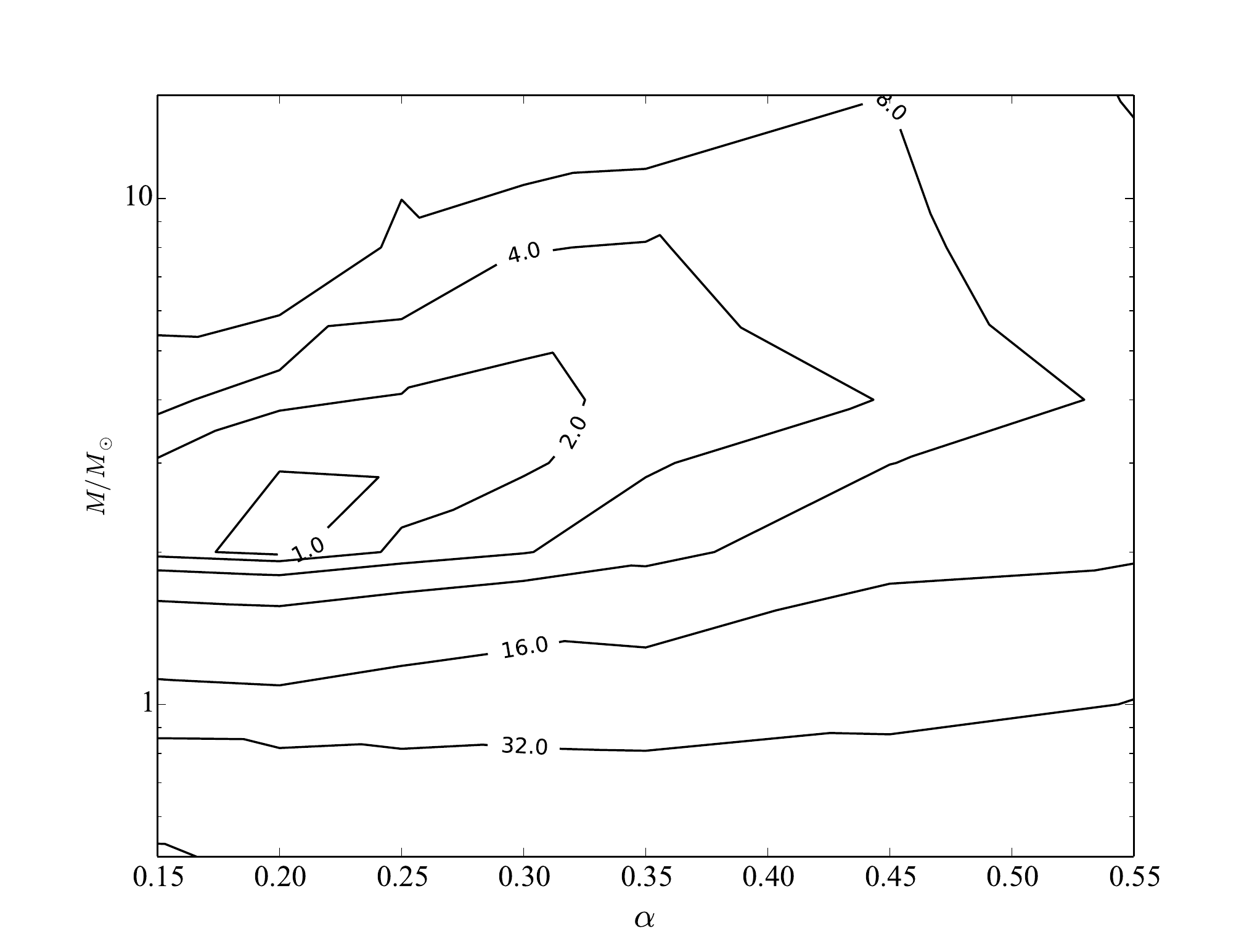}
\vskip -0 truecm
\caption{{Reduced $\chi^2(\alpha,M)$ corresponding to the comparison between  the microlensing magnification  {PDF}   of a bimodal distribution of stars and BHs ($M_*\sim 0.3M_\odot$, $M_{BH}\sim 30 M_\odot$, $\alpha_*=\alpha_{BH}=0.2$),  and the microlensing magnification PDF of a single-mass distribution of mass fraction $\alpha$ and mass $M$ (see text).}  The size of the source is $r_s= 5\,\rm lt-day$.}
\end{figure}
\begin{figure}[h]
\vskip -1 truecm
\plotone{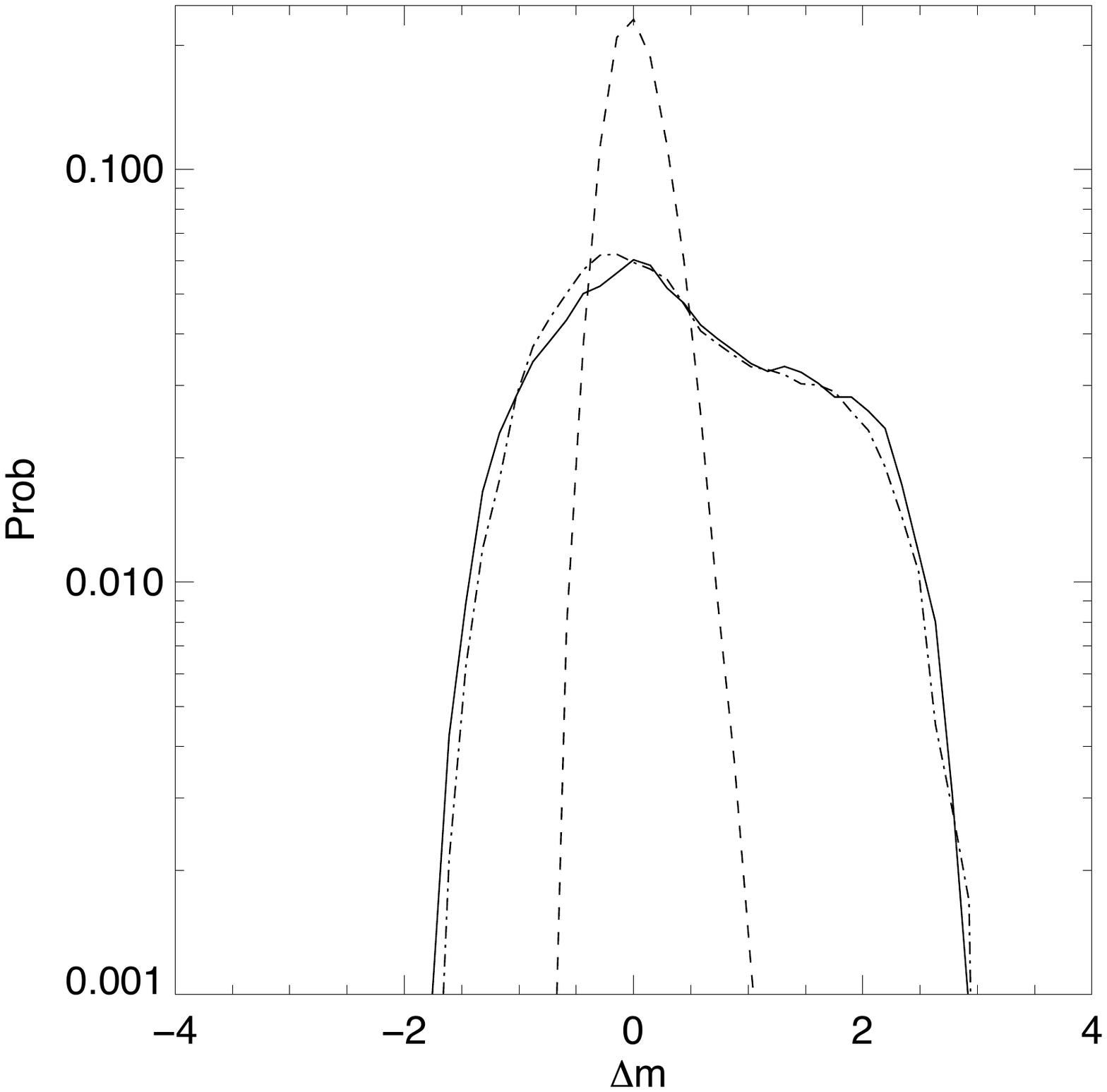}
\vskip -4 truecm
\caption{Comparison of the microlensing magnification  {PDF} corresponding to a bimodal distribution (thick continuous line) of stars and BH ($M_*\sim 0.3M_\odot$, $M_{BH}\sim 30 M_\odot$, $\alpha_*=\alpha_{BH}=0.2$) with the best fit ({dot-dashed} line) of a single-mass distribution (corresponding to $\alpha=0.2$, $M=2 M_\odot$) and with the fit (dashed line) corresponding to a single-mass distribution with a typical stellar mass ($\alpha=0.2$, $M=0.3 M_\odot$). The size of the source is $r_s= 5\,\rm lt-day$.}
\end{figure}
%
%
%

\clearpage


\end{document}